\documentclass[english]{revtex4-1}
\usepackage[T1]{fontenc}
\usepackage[utf8]{luainputenc}
\usepackage{babel}
\usepackage{amssymb}
\usepackage{graphicx}
\usepackage{esint}
\usepackage[unicode=true,pdfusetitle,
 bookmarks=true,bookmarksnumbered=false,bookmarksopen=false,
 breaklinks=false,pdfborder={0 0 1},backref=false,colorlinks=false]
 {hyperref}

\makeatletter

\providecommand{\tabularnewline}{\\}

\makeatother

\begin{document}

\title{Random Polymers and Generalized Urn Processes}

\author{Simone Franchini and Riccardo Balzan}

\address{Sapienza Università di Roma, 1 Piazza Aldo Moro, 00185 Roma, Italy}
\begin{abstract}
We describe a microcanonical approach for polymer models that combines
atmospheric methods with urn theory. We show that Large Deviation
Properties of urn models can provide quite deep mathematical insight
by analyzing the Random Walk Range problem in $\mathbb{Z}^{d}$. We
also provide a new mean field theory for the Range Problem that is
exactly solvable by analogy with the Bagchi-Pal urn model.
\end{abstract}
\maketitle

\section{Introduction}

In this paper we present a novel approach to deal with microcanonical
polymer models derived in analogy with urn process theory. The main
point in this method is that it is possible to relate the density
of states of an interacting chain with the problem of computing the
large deviation behavior of an associated Markov urn process \cite{Johnson_Koz,MahmoudBook,Pemantle}
once the urn function of the problem is identified (see below). Here
we deal with models that can be related to a two-color urn, first
introduced by Hill, Lane and Sudderth \cite{Hill Lane Sudderth,Hill Lane Sudderth 2,Fajolet-Analytic Urns,FranchiniSPA},
for which a detailed Large deviations theory has been recently developed
\cite{FranchiniSPA}. In particular, we will provide an explicit example
by studying the classic Random Walk range problem (RP, \cite{Huges,Spitzer,Feller,Stanley,Franchini}),
that is, computing the number of different lattice sites visited by
a random walk of given lenght. 

The HLS urns is a Markov process first introduced in \cite{Hill Lane Sudderth}.
Consider an infinite capacity urn with a finite number of black and
white balls and let $y_{t}$ the fraction of black balls inside the
urn at a certain time $t$ of the evolution, then in a HLS process
of \textit{urn function} $\pi\left(y\right)$ at each step a black
ball is added with probability $\pi$$\left(y_{t}\right)$ and a white
one is added otherwise. The process is then parametrized by the function
$\pi\left(y\right)$, that represent the probability of adding a black
ball at the considered step conditioned that the urn has reached a
certain fraction of black balls. 

The second ingredient is the \emph{endpoint atmosphere}, introduced
some years ago within the study of the Self-Avoiding Walk as the number
of ways in which a chain of $N$ steps can be continued by adding
one monomer to the endpoint \cite{Rechnitzer}. As we shall see, it
is possible to combine this two ideas together and define HLS processes
that converge to a given polymer model in the thermodynamic limit
by interpreting the probability that adding a step to a given chain
produces an increase in energy analog to adding a black ball in the
the associated urn process, ie the number of black balls will represent
the total energy of our polymer. 

Before starting let introduce some notation. Let $\mathbb{L}$ be
some regular lattice and let $L_{1}$ be the possible orientations
on $\mathbb{L}$. Then we call a chain $\omega_{N}\in L_{1}^{N}$
of $N$ steps on $\mathbb{L}$ the ordered -sequence of steps $\delta x_{t}\in L_{1}$
for $1\leq t\leq N$ , with $L_{1}^{N}$ the set of distinct random
walks of $N$ steps on $\mathbb{L}$, thus $\omega_{N}=\{\delta x_{1},\,...\,,\delta x_{N}\}.$
If we fix the starting point $x_{0}$ we can also represent $\omega_{N}$
by the positions $x_{t}\in\mathbb{L}$, related to the steps $\delta x_{t}$
by $\delta x_{t}=x_{t}-x_{t-1}$. Hereafter we will assume that $x_{0}\equiv0$
and 
\begin{equation}
\omega_{N}=\{x_{0},x_{1},\,...\,,x_{N}\}
\end{equation}
with steps $x_{t}-x_{t-1}\in L_{1}$ for all times $1\leq t\leq N$. 

Now consider the interaction energy $H(\omega_{N})$, that is the
energy associated to the chain configuration $\omega_{N}$. We assume
that the interaction energy $H$ can be defined for arbitrary size
$N$ of the walks. In general, we can define the free energy density
per monomer of the interaction $H$ supported by $L_{1}^{N}$ in the
thermodynamic limit 
\begin{equation}
f\left(\beta\right)=-\lim_{N\rightarrow\infty}{\textstyle {\textstyle \frac{1}{\beta N}}}\log\,{\textstyle \sum_{\,\omega_{N}\in L_{1}^{N}}}e^{-\beta H\left(\omega_{N}\right)}.
\end{equation}
After rescaling by the number of possible walks we can write $-\beta f\left(\beta\right)=\log|L_{1}|+\zeta\left(\beta\right)$
where $\zeta\left(\beta\right)$ is the Cumulant Generating Function
(CGF) of the variable $H\left(\omega_{N}\right)$ 
\begin{equation}
\zeta\left(\beta\right)=\lim_{N\rightarrow\infty}{\textstyle \frac{1}{N}}\log\,\langle\,e^{-\beta H\left(\omega_{N}\right)}\rangle_{L_{1}^{N}}
\end{equation}
with average over $\omega_{N}$ taken uniform on $L_{1}^{N}$. 

Then, let $\omega_{N}$ be a random chain of $N$ step and define
the sequence $\omega_{t}\subset\omega_{N}$ subwalks of $\omega_{N}$
according to the monomer ordering $t$, ie $\omega_{t}=\{x_{0},\,...\,,x_{t}\}$.
In this paper we will deal with energy functions that satisfy 
\begin{equation}
H(\omega_{t+1})-H(\omega_{t})\in\left\{ 0,1\right\} 
\end{equation}
for all $\omega_{N}\in L_{1}^{N}$ and all $t$. This condition ensures
that the energy can either increase of one unit or do not increase
at all when a monomer is added to the endpoint of $\omega_{N}$, and
is an important technical point to connect with the HLS urns as it
allows to directly identify an increase in energy followed by one
step grows with adding a black ball to the associated urn. It is possible
to generalize to include more general transition spectra (multicolor
urns) but here we consider the binary cases as the LDP for such urns
have been already developed in detail \cite{FranchiniSPA}.

\section{HLS urns}

Before going further we need to introduce the HLS process \cite{Hill Lane Sudderth,Hill Lane Sudderth 2,Fajolet-Analytic Urns,FranchiniSPA}
and sketch some of its main properties we will use in the following.
An HLS urn is a two color urn process that is governed by a functional
parameter $\pi\left(y\right)$ called \textit{urn function} \cite{FranchiniSPA}.
Let us consider an infinite capacity urn containing two kinds of elements,
say black and white balls, and denote by 
\begin{equation}
Y=\left\{ \,Y_{t_{0}},Y_{t_{0}+1},\,...,Y_{N}\right\} ,
\end{equation}
the process describing the number of black balls inside the urn during
its evolution from $t=t_{0}$ to $N$. The process $Y$ evolves as
follows, let $y_{t}=Y_{t}/t$ be the fraction of black balls at time
$t$, then at step $t+1$ a new ball is added, whose color is black
with probability $\pi\left(y_{t}\right)$ and white with probability
$1-\pi\left(y_{t}\right)$.

Then, let $Y$ be an HLS urn process stopped at $N$, with initial
condition $Y_{t_{0}}=M_{0}$, describing the number of black balls
in the evolution of a HLS urn of urn function $\pi$. By simple arguments
on conditional expectations it is not hard to prove that the process
satisfy the following master equation 
\begin{equation}
P\left(Y_{N+1}=M+1\right)=\pi\left({\textstyle \frac{M}{N}}\right)\,P\left(Y_{N}=M\right)+\left(1-\pi\left({\textstyle \frac{M+1}{N}}\right)\right)\,P\left(Y_{N}=M+1\right)\label{eq:rrrr-1}
\end{equation}
that can be iterated backward to the initial condition 
\begin{equation}
P(Y_{t_{0}}=M_{0})=I(Y_{t_{0}}=M_{0}),
\end{equation}
where $\pi$ is the urn function and $I(Y_{t_{0}}=M_{0})$ is indicator
function, valued one if $Y_{t_{0}}=M_{0}$ and zero otherwise. 

In \cite{FranchiniSPA} the cumulant generating function of the process
\begin{equation}
\zeta\left(\beta\right)=\lim_{N\rightarrow\infty}{\textstyle \frac{1}{N}}\log\,{\textstyle \sum_{k\leq N}}\,e^{-\beta k}\,P\left(Y_{N}=k\right)
\end{equation}
is studied in detail and it is proven that it must satisfy the following
nonlinear differential equation 
\begin{equation}
\partial_{\beta}\,\zeta\left(\beta\right)=\pi^{-1}\left({\textstyle \frac{e^{\,\zeta\left(\beta\right)}-1}{e^{\,\beta}-1}}\right)
\end{equation}
with $\pi^{-1}$ inverse urn function. Of special interest for our
scopes will be the case of linear urn functions 
\begin{equation}
\pi\left(y\right)=a+by
\end{equation}
that in \cite{FranchiniSPA} are shown to be equivalent to the Baghi-Pal
model \cite{MahmoudBook,Fajolet-Analytic Urns}, a widely investigated
model due to its relevance in studying branching phenomena and random
trees (see \cite{MahmoudBook,Pemantle,Johnson_Koz} for some reviews).
Linear urn functions satisfy the differential equation
\begin{equation}
\partial_{\beta}\,\zeta\left(\beta\right)=-{\textstyle \frac{a}{b}}+{\textstyle \frac{1}{b}}\left({\textstyle \frac{e^{\,\zeta\left(\beta\right)}-1}{e^{\,\beta}-1}}\right).
\end{equation}
The above equation can be integrated exactly. Although the solution
depends on the considered parameter region, for our analysis it will
suffice to take $a>0$, $a+b<1$ and $b>0$, $\beta>0$. From Corollary
10 of \cite{FranchiniSPA} we have that
\begin{equation}
1-e^{-\zeta\left(\beta\right)}={\textstyle \frac{a}{b}}e^{\frac{a}{b}\beta}\left({\textstyle 1-e^{-\beta}}\right)^{\frac{1}{b}}B\left({\textstyle {\textstyle \frac{a}{b}},\frac{b-1}{b};1-e^{-\beta},1}\right)\label{eq:fff-1}
\end{equation}
where $B\left(q,p;u,v\right)$ is a Generalized Hypergeometric function
of the second kind
\begin{equation}
B\left(q,p;u,v\right)=\int_{u}^{v}dt\,\left(1-t\right)^{q-1}t^{p-1}.
\end{equation}
As we shall see in short our mean field theory will be described by
linear urn theory above.
\begin{figure}
\includegraphics[scale=0.3]{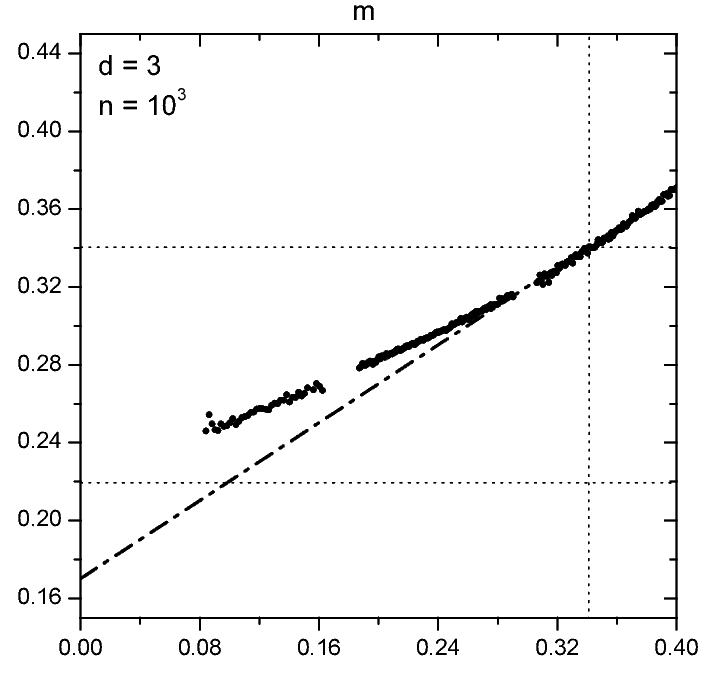}\caption{Numerical estimate of $\pi_{N}\left(\left\lfloor Nm\right\rfloor \right)$
from Eq. (\ref{eq:atmosfera}), for dimensions $d=3$ and $0\leq m\leq0.4$.
The lengths of the chains were $N=1000$. The vertical dot line and
the upper horizontal dot line show the Polya RW constant $C_{3}$
\cite{Douglas}, and crosses at the RW point. The lower horizontal
dotted line is the limit $\pi_{\infty}\left(0\right)=\eta_{3}$ \cite{RW and SAW theory,Madras-Slade}
according to \cite{Clisby}. The dash dotted line is the MF aproximation
$\pi_{MF}\left(m\right)=(C_{3}+m)/2$, computed according to Eq. (\ref{eq:28})
and (\ref{eq:29}).}
\end{figure}

\section{Urn analogy}

Although the limitations imposed by Eq. (4), simple two colors HLS
urns still allows to describe interesting models (that are not limited
to polymer physics). The problem we investigate here is the the Random
Walk Range problem (RP) on the cubic lattice $\mathbb{Z}^{d}$ \cite{Huges,Feller,Franchini},
a model showing \cite{Franchini} a full crossover from Self-Avoiding
Walks (SAW) \cite{Madras-Slade} to collapsed globular configuration
in the range density per monomer, and is shown to have an interesting
geometric \textit{Coil-to-Globule} transition (CG, the chain collapses
from an extended random coil to a liquid-like cluster, \cite{De Gennes,Flory})
at a critical range density for any $d\geq3$. 

Take a walk $\omega_{N}\in L_{1}^{N}$ and define the number of different
sites of $\mathbb{L}$ visited by $\omega_{N}$. We will approach
the RP by studying interaction energy 
\begin{equation}
H(\omega_{N})=N-R(\omega_{N})=N-{\textstyle \sum_{\,x\in\mathbb{Z}^{d}}\,}I\left(x\in\omega_{N}\right)
\end{equation}
an Hamiltonian first introduced by Stanley et Al. in \cite{Franchini-2,Stanley}.
To show the urn process analogy we first need to introduce some microcanonical
estimators. Let 
\begin{equation}
L_{N}\left(M\right)=\left\{ \,\omega_{N}\in L_{1}^{N}:\,H\left(\omega_{N}\right)=M\right\} 
\end{equation}
be the fraction of walks of length $N$ with an energy of exactly
$M$, then call $P\left(H(\omega_{N})=M\right)$ the probability that
a chain $\omega_{N}$ uniformly picked from $L_{N}$ has energy $M$.
Notice that the constraint of binary energy increase guarantees that,
for all these functions, $m$ is a real parameter between zero and
one. Then, let consider a walk of $N$ steps $\omega_{N}\in L_{1}^{N}$and
define the average of the energy after a random continuation $\omega_{1}^{*}\in L_{1}$
from the endpoint of $\omega_{N}$ 
\begin{equation}
\delta H_{1}\left(\omega_{N}\right)=\langle H(\omega_{N}\cup\omega_{1}^{*})-H(\omega_{N})\rangle_{L_{1}}=\frac{1}{|L_{1}|}{\textstyle \sum_{\,x\in\mathbb{Z}^{d}}\,}I\left(x\in\omega_{N}\right)I\,(x\in L_{1})\label{eq:delta}
\end{equation}
Since energy can increase only by zero or one then the average increase
$\delta H_{1}\left(\omega_{N}\right)$ equals the probability that
a random continuation of the walk $\omega_{N}$ from its endpoint
$x_{N}$ produces a self interaction according to $H$. We then define
the atmosphere 
\begin{equation}
\pi_{N}\left(M\right)=\langle\delta H_{1}\left(\omega_{N}\right)\rangle_{L_{N}\left(M\right)},\label{eq:atmosfera}
\end{equation}
that is the probability of self-intersection after a random continuation
of $\omega_{N}$, conditioned to the event that the range is $R(\omega_{N})=N-M$. 

It can be proven that $P\left(H(\omega_{N})=M\right)$ satisfy the
following Master equation
\begin{equation}
P\left(H(\omega_{N+1})=M\right)=\pi_{N}\left(M\right)P\left(H(\omega_{N})=M\right)+\left(1-\pi_{N}\left(M+1\right)\right)P\left(H(\omega_{N})=M+1\right)\label{eq:2}
\end{equation}
 with initial condition
\begin{equation}
P\left(H(\omega_{1})=M\right)=I\left(M=0\right).
\end{equation}
If we take $H(\omega_{N})=Y_{N}$ it is clear that the Master Equation
for the measure of the event is the same of the event $Y_{N}=M$ of
an HLS urn of non-homogeneous urn function $\pi_{N}\left(M\right)$.
In \cite{FranchiniSPA} it is shown that if
\begin{equation}
\lim_{N\rightarrow\infty}\left|\,\pi_{N}(\left\lfloor Nm\right\rfloor )-\pi\left(m\right)\right|=0
\end{equation}
then the cumulant generating function of the process is the same of
an HLS urn of urn function $\pi\left(m\right)$. The existence of
$\pi\left(m\right)$ for the RP can be inferred by subadditivity,
but we do not give a proof here because the convergence of $\pi_{N}(\left\lfloor Nm\right\rfloor )$
toward some smooth $\pi\left(m\right)$ is already clear from our
numerical analysis (see Figure 3).
\begin{figure}
\includegraphics[scale=0.3]{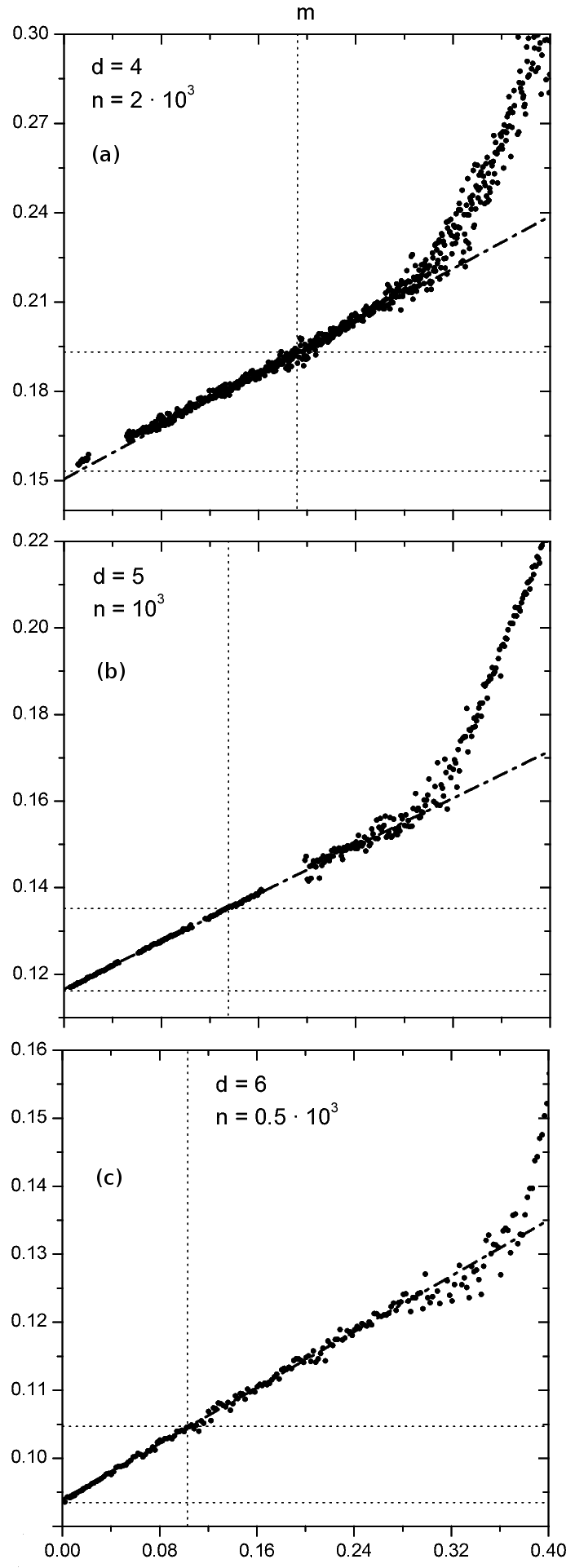}\caption{Numerical estimation of $\pi_{N}\left(\left\lfloor Nm\right\rfloor \right)$
from Eq. (\ref{eq:atmosfera}), for $d=4,5,6$ and $0\leq m\leq0.4$.
As for $d=3$, the lengths of the chains were $N=2000$ for $d=4$,
$N=1000$ for $d=5$ and $N=500$ for $d=6$. For each plot, the vertical
and the upper horizontal dotted lines are Polya constants $C_{d}$
\cite{Douglas,Huges}, while the lower horizontal dot lines are $\pi_{\infty}\left(0\right)=\eta_{d}$
from \cite{RW and SAW theory,Madras-Slade}. The dash dotted lines
are $\pi_{MF}\left(m\right)=C_{d}\left(1-B_{d}\right)+B_{d}\,m$ of
Eq. (\ref{eq:28}) and (\ref{eq:29}) (not linear fits).}
\end{figure}
 
\begin{figure}
\includegraphics[scale=0.3]{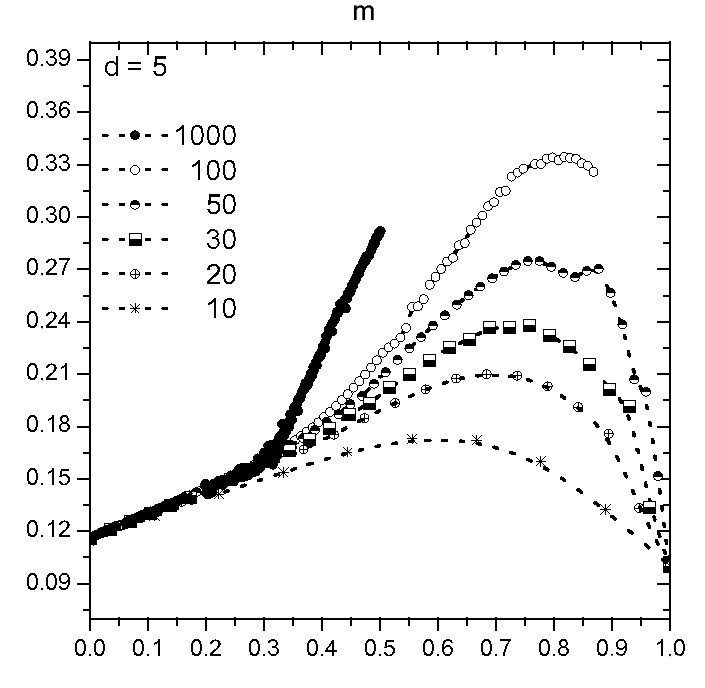}\caption{Numerical estimation of $\pi_{N}\left(\left\lfloor Nm\right\rfloor \right)$
from Eq. (\ref{eq:atmosfera}) for $d=5$, $0\leq m\leq1$ and from
$N=10$ to $N=1000$. The convergence to some limit urn function $\pi\left(m\right)$
is observed in the SAW region $m<m_{c}$.}
\end{figure}

\section{Numerical results}

In Figures 1,2 and 3 we present our numerical results concerning the
urn function $\pi_{N}\left(M\right)$ associated to the RP on $\mathbb{Z}^{d}$,
$3\leq d\leq6$, for which some properties can be deduced also from
known results in Random and Self-Avoiding Walks theory \cite{RW and SAW theory,Madras-Slade,Huges}. 

The numerical simulations where performed by a standard implementation
of the Pruned-Enriched Rosenbluth Method, PERM, see \cite{PERM,Grassberger,Prellberg,Hsu-Grassberger}.
For $3\leq d\leq6$ we restricted our attention to the region $M/N<m_{c}$,
where the typical configuration of $\omega_{N}$ is supposed to be
in the universality class of the self-avoiding walk. 

In a previous paper \cite{Franchini} we numerically studied the event
$H(\omega_{N})=\left\lfloor mN\right\rfloor $ and found a CG transition
for some critical value $m=m_{c}\in\left(0,1\right)$. We studied
the critical exponent governing the mean square displacement, 
\begin{equation}
\nu_{d}\left(m\right)=\lim_{N\rightarrow\infty}\frac{\log\,\langle x_{N}^{2}\rangle_{C_{N}\left(\left\lfloor Nm\right\rfloor \right)}}{2\log N},
\end{equation}
concluding the above limit exists and $\nu_{d}\left(m<m_{c}\right)=\nu_{d}$,
$\nu_{d}\left(m=m_{c}\right)=\nu_{c}$, $\nu_{d}\left(m>m_{c}\right)=1/d$,
where $\nu_{d}$ is the critical exponent governing the end-to-end
distance of the Self-Avoiding Walk \cite{Madras-Slade}. Also, $\nu_{c}=1/2$,
$m_{c}=C_{d}$ Polya constant \cite{Douglas} for $d=3,4$, and that
for $d\geq5$ it is expected that $m_{c}>C_{d}$ (see \cite{Franchini}
for further details about this topics). 

Here we observe that $\pi_{N}\left(M\right)$ approaches to some continuous
$\pi\left(M/N\right)$ uniformly on the considered range. Quite surprisingly,
we also observe that for $d\geq4$ the function $\pi$ suddenly approaches
some linear function (see Figures 2 and 3)
\begin{equation}
\pi_{MF}\left(m\right)=a_{RP}+b_{RP}\,m
\end{equation}
in the region $0\leq m\leq m_{c}$. Assuming a linear urn function,
the coefficients can be computed exactly from RW theory by relating
them to the variance of the energy $\sigma_{d}^{2}$. 

The constant $\sigma_{d}$ can be computed from Jain-Pruitt theorem
on the variance of the RP (\cite{Huges,Jian-Pruitt,Jiain-Orey,Dvoretzky-Erdos,Torney},
see also \cite{Den Hollander} for an explicit computation). For $d=3$,
Jain and Pruitt have shown \cite{Huges,Jian-Pruitt,Jiain-Orey} that
the leading order of the variance of $R(\omega_{N})$ for a random
walk is $\sigma_{3}^{2}N\log\left(N\right)$ with $\sigma_{3}$ exactly
computable, while for for $d\geq4$ the same authors shows that the
variance is $\sigma_{d\,}^{2}N$ with $\sigma_{d}$ expressed by the
relation 
\begin{equation}
\sigma_{d}^{2}=C_{d}(1-C_{d})+2a.
\end{equation}
Accurate estimates for $C_{d}$ are in \cite{Douglas}. To determine
$a$ we follow \cite{Huges,RW and SAW theory,Jiain-Orey,Jian-Pruitt,Den Hollander}.
Let first introduce the propagator 
\begin{equation}
G\left(x\right)=\int_{\left[-\pi,\pi\right]^{d}}{\textstyle \frac{dq}{\left(2\pi\right)^{d}}\,}e^{iqx}(1-\tilde{\lambda}_{d}\left(q\right))^{-1},\label{eq:2.0.7}
\end{equation}
where $\tilde{\lambda}_{d}\left(q\right)$, $q\in\left[-\pi,\pi\right]^{d}$
is the \emph{structure factor} of the hypercubic lattice $\mathbb{Z}^{d}$
\begin{equation}
{\textstyle \tilde{\lambda}_{d}\left(q\right)=\frac{1}{d}\sum_{i=1}^{d}\cos\left(q_{i}\right),}
\end{equation}
and where $q_{i}$ are the components of the dual vector $q$. The
quantity $G\left(x\right)$ represents the expected number of visits
to a given site $x\in\mathbb{Z}^{d}$ for an infinite lenght random
walk. From standard random walks theory follows \cite{Huges} 
\begin{equation}
{\textstyle \frac{C_{d}}{1-C_{d}}}={\textstyle \sum_{\,x\,\in\,\mathbb{Z}^{d}\setminus\left\{ 0\right\} }}G\left(x\right)
\end{equation}
 for the Polya constants. For $d\geq3$, by Jain-Pruitt Theorem it
is also possible to write $a$ in terms of the $G\left(x\right)$
function as well \cite{Den Hollander}: 
\begin{equation}
{\textstyle a=}\,{\textstyle \sum_{\,x\,\in\,\mathbb{Z}^{d}\setminus\left\{ 0\right\} }}\ {\textstyle \frac{(1-C_{d})^{4}\,G\left(x\right)^{3}}{1+(1-C_{d})\,G\left(x\right)}}\label{eq:2.11}
\end{equation}
Then, from the convergence condition of a generic HLS urn $C_{d}=\pi\left(C_{d}\right)$
(see \cite{FranchiniSPA}) follows that 
\begin{equation}
a_{RP}=C_{d}\left(1-b_{RP}\right).\label{eq:28}
\end{equation}
By computing the variance of the linear urn from the CGF of Eq. (11)
and confronting with the expression of the RP variance from Jain-Pruitt
Theorem above we get 
\begin{equation}
{\textstyle b_{RP}=\frac{1}{2}\left({\textstyle 1-\frac{C_{d}(1-C_{d})}{\sigma_{d}^{2}}}\right).}\label{eq:29}
\end{equation}
Linear urns with the above values are shown as dot lines in Figures
1 and 2. A detailed computation will be presented elsewhere.

\section{Two colors Mean-field theory }

Besides the computational advantages in numerically studying the atmosphere
instead of counting the number of walks, that has been already exploited
in \cite{Rechnitzer}, the urn theory allows for new interesting analytic
approaches. For example, here we give a simple model that match the
linear urn theory suggested by our numerical simulations. In the spirit
of the classic Pincus-De Gennes blob picture \cite{De Gennes} let
slice the chain $\omega_{N}$ into a number $n$ of sub-chains
\begin{equation}
\omega_{N}=\left\{ \omega_{T}^{0},\omega_{T}^{1},\,...\,,\omega_{T}^{n}\right\} 
\end{equation}
each of size $T=N/n$. The sub-chains are indicated with 
\begin{equation}
\omega_{T}^{i}=\left\{ x_{0}^{i},\,x_{1}^{i},\,...\,,x_{T}^{i}\right\} \subset\omega_{N}
\end{equation}
and satisfy the chain constraint 
\begin{equation}
x_{T}^{i}=x_{0}^{i+1}.
\end{equation}
If we neglect the mutual self-intersections between different blocks
we can approximate the energy with 
\begin{equation}
H\left(\omega_{N}\right)\simeq{\textstyle \sum_{i=1}^{n}H\left(\omega_{T}^{i}\right)},\label{eq:33}
\end{equation}
and the energy increment
\begin{equation}
\delta H_{1}\left(\omega_{N}\right)\simeq\delta H_{1}\left(\omega_{T}^{1}\right).\label{eq:34}
\end{equation}
The probability measure conditioned to $H\left(\omega_{N}\right)=\left\lfloor Nm\right\rfloor $
is then approximated by a product measure
\begin{equation}
\mu_{m}\left(\omega_{N}\right)\simeq{\textstyle \prod_{i=1}^{n}}\,\mu_{m_{i}}\left(\omega_{N}^{i}\right).
\end{equation}

Notice that the approximation of Eq.s (\ref{eq:33}), (\ref{eq:34})
is expected to hold at least if both $N,T\rightarrow\infty$ and $d\geq4$
because above the critical dimension the interaction between different
subwalks is negligible in the thermodynamic limit \cite{Madras-Slade,Spitzer,Feller,Huges}.
\begin{table}
\centering{}%
\begin{tabular}{|c|c||c|c|c|}
\hline 
\noalign{\vskip\doublerulesep}
$d$ & $b_{RP}$ & $\eta_{d}^{(u)}$ & $\eta_{d}$ & $\eta_{p}^{\left(u\right)}/\eta_{d}-1$\tabularnewline[\doublerulesep]
\hline 
\hline 
$3$ & $1/2$ & $0.17026\left(9\right)$ & $0.2193\left(5\right)$ & $-22.38\%$\tabularnewline
\hline 
$4$ & $0.22080\left(9\right)$ & $0.15054\left(1\right)$ & $0.1532445\left(6\right)$ & $-1.76\%$\tabularnewline
\hline 
$5$ & $0.13767\left(2\right)$ & $0.11656\left(8\right)$ & $0.1161456\left(3\right)$ & $0.36\%$\tabularnewline
\hline 
$6$ & $0.10266\left(0\right)$ & $0.09396\left(5\right)$ & $0.0934921\left(3\right)$ & $0.51\%$\tabularnewline
\hline 
$7$ & $0.08291\left(2\right)$ & $0.078727\left(3\right)$ & $0.07837021\left(4\right)$ & $0.46\%$\tabularnewline
\hline 
$8$ & $0.07030\left(6\right)$ & $0.067786\left(5\right)$ & $0.0675464\left(2\right)$ & $0.36\%$\tabularnewline
\hline 
\end{tabular}\\
\caption{In table is shown $\eta_{d}^{(u)}=C_{d}(1-b_{RP})$ from MF theory
and numerically determined $\eta_{d}$ from literature \cite{Clisby,Clisby2,Owczarek-Prellberg-1,Madras-Slade}:
while $d=3$ there is a heavy underestimation (more that $20\%$),
yet for $d=4$ there is error of $2\%$, and $d\geq5$ under the percent.
An exhaustive analysis of our results about the range problem will
be published in a dedicated paper.}
\end{table}
 If instead we take $T$ to be finite then the mutual intersections
between the segments are no longer negligible, nonetheless, since
the typical length between two self intersection is of order $O\left(1/2d\right)$
we expect that the above linear approximation will be asymptotically
exact also for $T<\infty$ in the limit $d\rightarrow\infty$. 

Now, we approximate by assuming that the sub-chains distributions
can be of two kind only, say $A$ and $B$ 
\begin{equation}
\mu_{i}\left(\omega_{N}^{i}\right)=\varphi^{i}\mu_{A}\left(\omega_{N}^{i}\right)+\left(1-\varphi^{i}\right)\mu_{B}\left(\omega_{N}^{i}\right).\label{eq:34-1}
\end{equation}
This recall again the two colors approximation, and seems a crucial
technical point to obtain linear urns. We can give a simple physical
understanding of this by taking $A$ to be, for example, a Self-Avoiding
Walk $m_{A}=0$, $T$ equal to the average number of steps a SRW can
do without self-intersectiong, and $B$ to contain a self-intersection
such that the local range density is $1-m_{B}$ with $m_{B}=C_{d}>0$.
Forcing a self-intersection in one block will certainly bring to a
decrease in the total range density, on the other side this will affect
the atmosphere only if the self-intersection happens near the the
endpoint where we are supposed to grow the chains. 

In the previous formula Eq. (\ref{eq:34-1}) we introduced a binary
sequence 
\begin{equation}
\varphi=\{\varphi^{1},...\,,\varphi^{n}\},
\end{equation}
with $\varphi^{i}\in\left\{ 0,1\right\} $, that keep record of weather
a subchain is either of one kind or the other, and can be interpreted
as the color of the ball we add. For a walk in a given state we assume
that the range density is peaked around some value 
\begin{equation}
H\left(\omega_{T}^{i}\right)\simeq m_{A}\,T\,\varphi^{i}+m_{B}\,T\left(1-\varphi^{i}\right)
\end{equation}
concerning the energy and
\begin{equation}
\delta H_{1}\left(\omega_{N}\right)\simeq\pi_{B}+\left(\pi_{A}-\pi_{B}\right)\varphi^{1}
\end{equation}
for the energy increment. Given this we find 
\begin{equation}
H\left(\omega_{N}\right)/N\simeq m_{B}+\left(m_{A}-m_{B}\right)\,{\textstyle \frac{1}{n}\sum_{i=1}^{n}\varphi^{i}},
\end{equation}
then, taking the average over $C_{N}\left(M\right)$ with
\begin{equation}
\langle\,\varphi^{1}\rangle_{C_{N}\left(M\right)}={\textstyle \frac{1}{n}\sum_{i=1}^{n}\langle\,\varphi^{i}\rangle_{C_{N}\left(M\right)}}
\end{equation}
we arrive to a linear expression for the urn function
\begin{equation}
\pi_{N}\left(M\right)\simeq{\textstyle a_{RP}+b_{RP}}\,M/N,
\end{equation}
with coefficients $a_{RP}=\pi_{B}-m_{B}\,b_{RP}$ and 
\begin{equation}
b_{RP}={\textstyle \frac{\pi_{A}-\pi_{B}}{m_{A}-m_{B}}}.
\end{equation}
There are various ways to obtain these coefficients from Random Walks
Theory. If we take $A$ to be the RW and $B$ to be the SAW, we arrive
to the linear urn described before, where $a_{RP}$ equals the SAW
normalized connective constant $\eta_{d}$ \cite{RW and SAW theory,Madras-Slade},
and $b_{RP}={\textstyle 1-\eta_{d}/C_{d}}$. By comparing to mean-field
value we obtain an expression for the rescaled connective constant
of the Self-Avoiding Walk \cite{RW and SAW theory,Madras-Slade}
\begin{equation}
\eta_{d}\simeq{\textstyle \frac{C_{d}}{2}\left(1+\frac{C_{d}(1-C_{d})}{\sigma_{d}^{2}}\right)=\eta_{d}^{(u)}},
\end{equation}
A computation of $\eta_{d}^{(u)}$ via numerical integration (see
Table I) suggests to exclude that this is the correct value for $\eta_{d}$,
at least for $d\leq8$, although our numerical analysis show narrow
discrepancies as $d$ increases. 

We conclude by remarking that in the above mean field theory a critical
ingredient is to assume that we can obtain the urn function via interpolation
between any two fixed energy states, for example sub-chains that are
either self-avoiding or critically collapsed, or between self avoiding
chain and random chains as well. The reason for this to hold so well
in high dimensions is not clear. 

The numerical estimates $\eta_{d}^{(u)}$ in Table 1, based on the
linear urn analogy and the Jain-Puitt theorem, seems to indicate that
Eq. (38) is slightly deviating from the accurate numerical values
available in literature, at least for $d\leq8$. Unfortunately, we
expect this simple linear urn analogy to be only asymptotic for $d\rightarrow\infty$,
but we also expect that more refined estimate of the urn function
can be obtained by a proper accounting of mutual self intersections
between the blocks. Further investigations on this aspect would be
of certain interest, also, it would be interesting to understand the
meaning of higher order polynomial urn functions. We expect that some
light on this may be obtained at least in $d=4$ by confronting with
a recently developed exact renormalization scheme based on Lace expansion
\cite{Slade}.

\section{Acknowledgments}

We would like to thank Giorgio Parisi (Sapienza Univeristà di Roma),
Jack F. Douglas (NIST), Pietro Caputo (Università Roma Tre) and Valerio
Paladino (Amadeus IT) for interesting discussions and suggestions.
This project has received funding from the European Research Council
(ERC) under the European Union’s Horizon 2020 research and innovation
programme (grant agreement No {[}694925{]}).


\begin{thebibliography}{10}
\bibitem{RW and SAW theory}It can be shown that $\pi\left(0\right)=\eta_{d}=1-\mu_{d}/2d$,
where $\mu_{d}$ is the connective constant of the Self-Avoiding Walk
\cite{Madras-Slade}. Also, it is possible to compute the first derivative
of $\pi$ for $m=C_{d}$ exactly \cite{Jiain-Orey,Jian-Pruitt}. This
topic will be treated in detail elsewhere.
\bibitem{Franchini-2}The Stanley Model (SM, see \cite{Stanley} for
the original formulation, or \cite{Huges} for a review) is a model
of correlated random walks which describes polymer chains with either
repulsion or attractions. The attractive case is mathematically equivalent
to the Rosenstock Trapping Model (random walk on lattice with randomly
distributed irreversible traps: a walker encountering a trap is killed
there, see \cite{Huges}).
\end{thebibliography}
\end{document}